\nonstopmode
\documentclass[12pt,a4paper]{article}
\usepackage{myprooftree}
\usepackage{html}

\newcommand{\foetus}{{\sf foetus}}
\newcommand{\lmu}{https://www.lmu.de/en/}
\newcommand{\inforoot}{https://www.ifi.uni-muenchen.de}
\newcommand{\informatik}{\inforoot/index.html}
\newcommand{\tcsroot}{https://www2.tcs.ifi.lmu.de}
\newcommand{\tcs}{\tcsroot/index.html}
\newcommand{\homepage}{\inforoot/~abel/}
\newcommand{\cgi}{\tcsroot/cgi-bin/foetus-cgi}
\newcommand{\myemail}{abel@informatik.uni-muenchen.de}
\newcommand{\foetusftp}{\homepage ftp/foetus/}

\newcommand{\htmllink}[2]{\html{\htmladdnormallink{#1}{#2}}}

\newcommand{\runcgi}[2]{\htmllink{#1}{\cgi ?DEFS=#2}}
\newcommand{\runexample}[1]{\runcgi{Run example.}{#1}}

\newcommand{\srcfile}[1]{\htmladdnormallink{{\tt #1}}{\foetusftp#1}}
\newcommand{\download}{SML sources: \url{https:github.com/andreasabel/foetus}}

\newcommand{\Pitype}[1]{\{#1\}}
\newcommand{\Let}[2]{\mathrm{Let}\ #1\ \mathrm{in}\ #2}
\newcommand{\xlet}[2]{\mathrm{let}\ #1\ \mathrm{in}\ #2}
\newcommand{\case}[2]{\mathrm{case}\ #1\ \mathrm{of}\ \{ #2 \}}
\newcommand{\darr}{\Rightarrow}
\newcommand{\G}{\Gamma}
\newcommand{\nat}{\mathrm{nat}}
\newcommand{\alist}{\mathrm{list}}
\newcommand{\bool}{\mathrm{bool}}
\newcommand{\ord}{\mathrm{ord}}

\newtheorem{example}{Example}[section]

\usepackage{amsbsy}
\usepackage{amsfonts}
\newcommand{\N}{\mathbb{N}} 
\newcommand{\Z}{\mathbb{Z}}
\newcommand{\less}{\boldsymbol{<}}
\newcommand{\equal}{\boldsymbol{=}}
\newcommand{\unknown}{\boldsymbol{?}}
\newcommand{\quotrel}[1]{\mathrm{`}\!#1\!\mathrm{'}}
\newcommand{\ttverb}[1]{{\tt #1}}
\newcommand{\tttilde}{$\mathtt{\sim}$}
\newcommand{\infixless}{\less}
\newcommand{\infixequal}{\equal}
\newcommand{\infixunknown}{\,\unknown\,}

\begin{htmlonly}
\newcommand{\N}{\mathbf{N}}
\newcommand{\Z}{\mathbf{Z}}
\newcommand{\less}{`\!<\!'} 
\newcommand{\equal}{`\!=\!'} 
\newcommand{\unknown}{`?'}
\newcommand{\quotrel}[1]{#1} 
\newcommand{\ttverb}[1]{{\tt #1}}
\newcommand{\tttilde}{\verb!~!}
\newcommand{\infixless}{\,\less\,}
\newcommand{\infixequal}{\,\equal\,}
\newcommand{\infixunknown}{\,\unknown\,}
\end{htmlonly}

\newcommand{\quotless}{\quotrel{\less}}
\newcommand{\quotequal}{\quotrel{\equal}}

\newcommand{\el}{\,\epsilon\,}

\newcommand{\finitsubset}{\subset_{\mathrm{finit}}}
\newcommand{\qed}{\hfill $q \cdot e \cdot d$}

\title{\foetus\ - Termination Checker for Simple Functional Programs}
\author{\htmladdnormallink{Andreas Abel}{\homepage}\thanks{
\htmladdnormallink{Theoretical Computer Science}{\tcs},
\htmladdnormallink{Institute of Computer Science}{\informatik},
\htmladdnormallink{Ludwigs-Maxi\-mi\-lians-University}{\lmu},
Oettingenstr.~67, D-80538~Munich, Germany,
email: \htmladdnormallink{{\tt \myemail}}{mailto:\myemail}.
I want to thank my supervisor \htmladdnormallink{Thorsten
  Altenkirch}{\tcsroot/~alti/} and \htmladdnormallink{Rolf
  Backofen}{\tcsroot/~backofen/} for his friendly support in technical
questions.
}}
\date{July 16, 1998}

\begin{document}
\maketitle


\begin{abstract}
We introduce a simple functional language \foetus\ (lambda calculus with
tuples, constructors and pattern matching) supplied with a termination
checker.
This checker tries to find a well-founded structural order on the
parameters on the given function to prove termination.
The components of the check algorithm are: function call extraction
out of the program text, call graph completion and finding a lexical
order for the function parameters.
\end{abstract}


\download

\section{Introduction}
Since the very beginning of informatics the problem of termination has
been of special interest, for it is part of the problem of program
verification for instance. Because the halting problem is
undecidable, there is no method that can prove or disprove termination
of all programs, but for several systems termination checkers have been
developed. We have focused on functional programs and designed the
simple language \foetus\footnote{In German \foetus\ is an abbreviation
  of ``Funktionale -- Obgleich Eingeschr\"ankt -- Termination
  Untersuchende Sprache'' {\bf ;-)}. It also expresses that it is
  derived from MuTTI (this is the German term for {\it Mum}).},
for which we have implemented a termination
prover. \foetus\ is a simplification of MuTTI (Munich Type Theory
Implementation) based on 
partial Type Theory (ala Martin L\"of) extended by tuples,
constructors and pattern matching. For the syntax see section
\ref{sec:syntax}.

To prove the termination of a functional program there has to be a
well founded order on the product of the function parameters
such that the arguments in each recursive call
are smaller than the corresponding input regarding this order.
We have limited to {\em structural} orderings.
\foetus\ tries to find such
an order by collecting all recursive calls of the given function and
the belonging behaviour of the function arguments. To handle mutually
recursive functions a call graph is constructed and completed.

Section \ref{sec:language} introduces the \foetus\ ``body'' (syntax and type
system). Section \ref{sec:examples} provides some examples to
intuitively learn the language and see the interpreter and termination
checker work. Then in section \ref{sec:outline} we explain the
``heart'' of \foetus: the call extractor; we also informally introduce
call graph completion and finding of the lexical order: the ``brain'' of
\foetus. The latter is formally described in section \ref{sec:formal}.

\section{\foetus\ Language\label{sec:language}}

\subsection{\foetus\ Syntax\label{sec:syntax}}

A \foetus\ program consists of {\em terms} and {\em definitions}.
\begin{center}
\begin{tabular}{rcll}
$P$ & $::$ &                       & {\it empty program} \\
    & $|$  & {\it term};       $P$ & {\it term to be evaluated} \\
    & $|$  & {\it definition}; $P$ & {\it definition for further use}
\end{tabular}
\end{center}
When processing input, \foetus\ evaluates the terms and stores the
definitions in the environment. ``Reserved words'' in the \foetus\
language are {\tt case}, {\tt of}, {\tt let} and {\tt in}.
Special characterss are {\tt ( ) [ ] \{ \} | .\ , ; = =>}. An identifier may
contain letters, digits, apostrophies and underscores.
If it starts with a small letter it stands for a variable,
else it denotes a constant.
\paragraph*{Term syntax.} In the following
$x$, $x_1$, $x_2$, $\dots$ denote variables,
$C$, $C_1$, $C_2$, $\dots$ constants and
$u$, $t$, $t_1$, $t_2$, $\dots$ \foetus\ terms.
\begin{center}
\begin{tabular}{rcll}
$t$ & $::$ & $x$       & {\it variable} \\
    & $|$  & $[x]t$    & {\it lambda} \\
    & $|$  & $t u$     & {\it application} \\
    & $|$  & $C(t)$    & {\it constructor} \\
    & $|$  & case $t$ of \{ $C_1 x_1 \darr t_1 | \dots | C_n x_n
      \darr t_n$ \}    & {\it pattern matching} \\
    & $|$  & $(C_1 = t_1, \dots, C_n = t_n)$
                       & {\it tuple} \\
    & $|$  & $t.C$     & {\it projection} \\
    & $|$  & let $x_1 = t_1, \dots, x_n = t_n$ in $t$
                       & {\it let} \\
    & $|$  & (t)       & {\it (extra parentheses) }
\end{tabular}
\end{center}

\paragraph*{Definitions.}
A definition statement has the form $x_1 = t_1, \dots, x_n = t_n$ (it
is a let-term without a ``body''). All variables $x_1, \dots, x_n$ are
defined simultanously, thus can refer to each other.

\paragraph*{Example.} The following \foetus\ program defines addition
on natural numbers (spanned by the two constructors {\tt O} ``zero'' and
{\tt S} ``successor'') and calculates $1+1$.

\begin{verbatim}
add = [x][y]case x of
        { O z => y
        | S x' => S(add x' y) };
one = S(O());
add one one;
\end{verbatim}

Note that although {\tt O} is a zero-argument-constructor the syntax
forces us to supply a dummy variable {\tt z} within the pattern
definition and also empty tuple {\tt ()} in the definition of {\tt one}.


\subsection{\foetus\ Type System}
In the following $x$, $x_1$, $x_2$, $\dots$ denote variables,
$C$, $C_1$, $C_2$, $\dots$ constants,
$u$, $t$, $t_1$, $t_2$, $\dots$ \foetus\ terms,
$\tau$, $\sigma$, $\sigma_1$, $\sigma_2$, $\dots$ \foetus\ types and
$X$, $X_1$, $X_2, \dots$ type variables.
$\G = x_1:\sigma_1, \dots, x_n:\sigma_n$ denotes the context.
The judgement
$$
    \G \vdash t:\sigma
$$
means ``in context $\G$ term $t$ is of type $\sigma$''.
\pagebreak

\paragraph*{Type formation.}\nopagebreak[4]
\begin{center}
\begin{tabular}{rcll}
$\tau$ & $::$ & $\sigma \rightarrow \tau$
           & {\it $\rightarrow$-type} \\
    & $|$  & $\{ C_1 : \sigma_1 | \dots | C_n : \sigma_n \}$
           & {\it labeled sum type} \\
    & $|$  & $( C_1 : \sigma_1 , \dots, C_n : \sigma_n )$
           & {\it labeled product type} \\
    & $|$  & $\Pitype{ X } \tau$
           & {\it polymorphic type} \\
    & $|$  & $\tau \sigma$
           & {\it instantiation of polymorphic type} \\
    & $|$  & $\Let{X_1 = \sigma_1, \dots, X_n = \sigma_n}{\tau}$
           & {\it recursive type}
\end{tabular}
\end{center}
In the formation of a recursive type with Let $X_i$ may only appear
strict positiv in $\sigma_i$. We define congruence on types $\cong$ as
the smallest congruence closed under
$$
    \Let{\vec{X} = \vec{\sigma}}{\tau} \cong \tau[X_1 := \Let{\vec{X} =
\vec{\sigma}}{X_1}; \dots; X_n := \Let{\vec{X} =
\vec{\sigma}}{X_n}]
$$
($\vec{X}=\vec{\sigma}$ abbreviates $X_1 = \sigma_1, \dots, X_n =
\sigma_n$). Thus we can substitute congruent types:
$$
\ru{\G\vdash t:\sigma \quad \sigma\cong\tau}
   {\G \vdash t: \tau}
$$
For ploymorphic types we have rules like in System F:
$$
\rux{\G \vdash t : \sigma \quad \mbox{$X$ not free type variable in $\G$}}
    {\G \vdash t : \{X\}\sigma}
    {poly-i}
$$
$$
\rux{\G \vdash t: \{X\}\sigma}
    {\G \vdash t: \sigma[X:=\tau]}
    {poly-e}
$$

\subsection{Typing rules for \foetus\ terms}

We here only briefly introduce the typing rules. For more detailed
explanation, read a book about type theorie, e.g. \cite{NPS90}.

\paragraph*{Lambda abstraction.}
$$
\rux{\G,x:\sigma \vdash t:\tau}
    {\G \vdash [x]t : \sigma \rightarrow \tau}
    {\rightarrow-i}
$$

\paragraph*{Application.}
$$
\rux{\G \vdash t:\sigma\rightarrow\tau \quad \G \vdash u:\sigma}
    {\G \vdash t u : \sigma}
    {\rightarrow-e}
$$

\paragraph*{Constructor.}
\[
\rux{\G\vdash t:\sigma_i}
    {\G\vdash C_i(t) : \{ C_1 : \sigma_1 | \dots | C_n : \sigma_n \}}
    {\{\}-i}
\]

\paragraph*{Pattern matching.}
\[
\rux{\G\vdash t: \{ C_1 : \sigma_1 | \dots | C_n : \sigma_n \}\qquad
     \G,x_i:\sigma_i \vdash u_i : \sigma \ \mbox{for all $1 \leq i
       \leq n$}}
    {\G\vdash \case{t}{C_1(x_1)\darr u_1 | \dots | C_n(x_n)\darr u_n}
      : \sigma}
    {\{\}-e}
\]

\paragraph*{Tupels.}
\[
\rux{\G\vdash t_i : \sigma_i\ \mbox{for all $1\leq i\leq n$}}
    {\G\vdash (C_1=t_1,\dots,C_n=t_n) : ( C_1 : \sigma_1 , \dots ,
  C_n : \sigma_n )}
    {()-i}
\]

\paragraph*{Projection.}
\[
\rux{\G\vdash t:( C_1 : \sigma_1 , \dots , C_n : \sigma_n )} {\G\vdash t.C_i
  : \sigma_i} {()-e}
\]

\paragraph*{Let.}
\[
\rux{\G,x_1:\sigma_1,\dots,x_n:\sigma_n \vdash t_i : \sigma_i \
  \mbox{for all $1\leq i\leq n$} \qquad \G,x_1:\sigma_1,\dots
  ,x_n:\sigma_n \vdash u : \tau} {\G\vdash \xlet{x_1=t_1;\dots
    ;x_n=t_n}{u}:\tau} {let}
\]
In \foetus\ type checking is not yet implemented and it is assummed
that all terms entered are well typed. Of course, only for well typed
terms the termination check produces valid results.
\paragraph*{Example.} The following well-known example for
non-termination passes the \foetus\ termination checker, but it is not
well typed.
\begin{verbatim}
f = [x]x x;
a = f f;
\end{verbatim}
\foetus\ output:\nopagebreak
\begin{verbatim}
f passes termination check
a passes termination check
\end{verbatim}
\runexample{f+\%3D\%5Bx\%5Dx+x\%3B\%0D\%0Aa+\%3D+f+f\%3B\%0D\%0A}


\section{Examples\label{sec:examples}}




\subsection{Addition and multiplication\label{ex:add}}
On the natural numbers
$$
    \nat := \Let{\nat = \{ \mathtt{O()}|\mathtt{S(}\nat\mathtt{)} \}}{\nat}
$$
we define $\mathtt{add},\mathtt{mult} : \nat \rightarrow \nat \rightarrow \nat$:
\begin{verbatim}
add = [x][y]case x of
        { O z => y
        | S x' => S(add x' y) };
mult = [x][y]case x of
        { O z => O z
        | S x' => (add y (mult x' y)) };
add (S(S(O()))) (S(O()));
mult (S(S(O()))) (S(S(S(O()))));
\end{verbatim}
{\bf\foetus\ output:}\nopagebreak[4]
\begin{verbatim}
< =: add -> add
add passes termination check by lexical order 0
< =: mult -> mult
mult passes termination check by lexical order 0
result: S(S(S(O())))
result: S(S(S(S(S(S(O()))))))
\end{verbatim}
\runexample{add+\%3D+\%5Bx\%5D\%5By\%5Dcase+x+of+\%0D\%0A\%09\%7B+O+z+\%3D\%3E+y\%0D\%0A\%09\%7C+S+x\%27+\%3D\%3E+S\%28add+x\%27+y\%29+\%7D\%3B\%0D\%0Amult+\%3D+\%5Bx\%5D\%5By\%5Dcase+x+of\%0D\%0A\%09\%7B+O+z+\%3D\%3E+O+z\%0D\%0A\%09\%7C+S+x\%27+\%3D\%3E+\%28add+y+\%28mult+x\%27+y\%29\%29+\%7D\%3B\%0D\%0Aadd+\%28S\%28S\%28O\%28\%29\%29\%29\%29+\%28S\%28O\%28\%29\%29\%29\%3B\%0D\%0Amult+\%28S\%28S\%28O\%28\%29\%29\%29\%29+\%28S\%28S\%28S\%28O\%28\%29\%29\%29\%29\%29\%3B\%0D\%0A}

\subsection{Subtraction\label{ex:sub}}
We define the predecessor function $\mathtt{p} : \nat \rightarrow
\nat$ and substraction on natural
numbers $\mathtt{sub} : \nat \rightarrow \nat \rightarrow \nat$. Note {\tt sub} $x$ $y$
calculates $y \stackrel{\cdot}{-} x$.
\begin{verbatim}
p = [x]case x of { O z => O z | S x' => x' };
sub = [x][y]case x of
        { O z => y
        | S x' => sub x' (p y) };
sub (S(S(O()))) (S(S(S(S(O())))));
\end{verbatim}
{\bf \foetus\ output:}\nopagebreak
\begin{verbatim}
p passes termination check
< ?: sub -> sub
sub passes termination check by lexical order 0
result: S(S(O()))
\end{verbatim}
\runexample{p+\%3D+\%5Bx\%5Dcase+x+of+\%7B+O+z+\%3D\%3E+O+z+\%7C+S+x\%27+\%3D\%3E+x\%27+\%7D\%3B\%0D\%0Asub+\%3D+\%5Bx\%5D\%5By\%5Dcase+x+of\%0D\%0A\%09\%7B+O+z+\%3D\%3E+y\%0D\%0A\%09\%7C+S+x\%27+\%3D\%3E+sub+x\%27+\%28p+y\%29+\%7D\%3B\%0D\%0Asub+\%28S\%28S\%28O\%28\%29\%29\%29\%29+\%28S\%28S\%28S\%28S\%28O\%28\%29\%29\%29\%29\%29\%29\%3B\%0D\%0A}

\subsection{Division\label{ex:div}}
Division
$\mathtt{div} : \nat \rightarrow \nat \rightarrow \nat$
can be implemented as follows in functional languages (note
that {\tt div x y} calculates $\lfloor \frac{y}{x} \rfloor$):
\begin{verbatim}
div (x,y) = div'(x,y+1-x)
div'(x,y) = if (y=0) then 0 else div'(x,y-x)
\end{verbatim}
{\tt div'} (just like division on natural numbers) terminates if the
divisor {\tt x} is unequal 0 because then {\tt y-x} $<$ {\tt y} in the
recursive call and thus one function argument is decreasing. But
\foetus\ recognizes only direct structural decrease and cannot see
that {\tt sub x y'} is less than {\tt y'}. To prove termination of
{\tt div'} you need a proof for $x \not= 0 \rightarrow \mathtt{sub}\, x \,y
< y$ \cite{BG96}.
\begin{verbatim}
p = [x]case x of { O z => O z | S x' => x' };
sub = [x][y]case x of
        { O z => y
        | S x' => sub x' (p y) };
div = [x][y]let
        div' = [y']case y' of
                { O z => O z
                | S dummy => S(div' (sub x y')) }
        in
        (div' (sub (p x) y));
div (S(S(O()))) (S(S(S(S(S(O()))))));
\end{verbatim}
{\bf\foetus\ output:}\nopagebreak
\begin{verbatim}
p passes termination check
< ?: sub -> sub
sub passes termination check by lexical order 0
div passes termination check
?: div' -> div'
div' FAILS termination check
result: S(S(O()))
\end{verbatim}
Here \foetus\ says {\tt div'} fails termination check, so {\tt div}
will not terminate either. {\tt div} would terminate, if {\tt div'}
terminated, therefore you get the answer {\tt div passes termination check}.
\runexample{p+\%3D+\%5Bx\%5Dcase+x+of+\%7B+O+z+\%3D\%3E+O+z+\%7C+S+x\%27+\%3D\%3E+x\%27+\%7D\%3B\%0D\%0Asub+\%3D+\%5Bx\%5D\%5By\%5Dcase+x+of\%0D\%0A\%09\%7B+O+z+\%3D\%3E+y\%0D\%0A\%09\%7C+S+x\%27+\%3D\%3E+sub+x\%27+\%28p+y\%29+\%7D\%3B\%0D\%0Adiv+\%3D+\%5Bx\%5D\%5By\%5Dlet+\%0D\%0A\%09div\%27+\%3D+\%5By\%27\%5Dcase+y\%27+of\%0D\%0A\%09\%09\%7B+O+z+\%3D\%3E+O+z\%0D\%0A\%09\%09\%7C+S+dummy+\%3D\%3E+S\%28div\%27+\%28sub+x+y\%27\%29\%29+\%7D\%0D\%0A\%09in\%0D\%0A\%09\%28div\%27+\%28sub+\%28p+x\%29+y\%29\%29\%3B\%0D\%0Adiv+\%28S\%28S\%28O\%28\%29\%29\%29\%29+\%28S\%28S\%28S\%28S\%28S\%28O\%28\%29\%29\%29\%29\%29\%29\%29\%3B\%0D\%0A}

\subsection{Ackermann function\label{ex:ack}}
The not primitive recursive Ackermann function
$\mathtt{ack} : \nat \rightarrow \nat \rightarrow \nat$.
\begin{verbatim}
ack = [x][y]case x of
        { O z  => S(y)
        | S x' => ack x' (case y of
                { O z  => S(O())
                | S y' => ack x y'} ) };
ack (S(S(O()))) (O());
\end{verbatim}
{\bf\foetus\ output:}\nopagebreak
\begin{verbatim}
foetus $Revision: 1.0 $
= <: ack -> ack
< ?: ack -> ack
ack passes termination check by lexical order 0 1
result: S(S(S(O())))
\end{verbatim}
\runexample{ack+\%3D+\%5Bx\%5D\%5By\%5Dcase+x+of\%0D\%0A\%09\%7B+O+z++\%3D\%3E+S\%28y\%29\%0D\%0A\%09\%7C+S+x\%27+\%3D\%3E+ack+x\%27+\%28case+y+of\%0D\%0A\%09\%09\%7B+O+z++\%3D\%3E+S\%28O\%28\%29\%29\%0D\%0A\%09\%09\%7C+S+y\%27+\%3D\%3E+ack+x+y\%27\%7D+\%29+\%7D\%3B\%0D\%0Aack+\%28S\%28S\%28O\%28\%29\%29\%29\%29+\%28O\%28\%29\%29\%3B\%0D\%0A}

\subsection{List processing\label{ex:list}}
We define lists over type $\alpha$ as
$$
    \alist := \Pitype{\alpha}\Let{ \alist = \{
      \mathtt{Nil()} | \mathtt{Cons(HD}:\alpha \mathtt{, TL}:\alist
      \mathtt{)} \} }{\alist}
$$
The well-known list processing functions $\mathtt{map} :  (\alpha
\rightarrow \beta) \rightarrow \alist \alpha \rightarrow \alist \beta$
and $\mathtt{foldl} : (\alpha \rightarrow \beta \rightarrow \beta)
\rightarrow \beta \rightarrow \alist \alpha \rightarrow \beta$ are
implemented and testet.
\begin{verbatim}
nil = Nil();
cons = [hd][tl]Cons(HD=hd,TL=tl);
l1 = cons (A()) (cons (B()) (cons (C()) nil));

map = [f][list]let
        map' = [l]case l of
                { Nil z => Nil()
                | Cons pair => Cons (HD=(f pair.HD),
                                     TL=(map' pair.TL))}
        in map' list;
map ([el]F(el)) l1;

foldl = [f][e][list]let
        foldl' = [e][l]case l of
                { Nil z => e
                | Cons p => foldl' (f p.HD e) p.TL }
        in foldl' e list;

rev = [list]foldl cons nil list;
rev l1;
\end{verbatim}
{\bf\foetus\ output:}\nopagebreak
\begin{verbatim}
nil passes termination check
cons passes termination check
l1 passes termination check

map passes termination check
<: map' -> map'
map' passes termination check by lexical order 0
result: Cons(HD=F(A()), TL=Cons(HD=F(B()), TL=Cons(HD=F(C()),
        TL=Nil())))

foldl passes termination check
? <: foldl' -> foldl'
foldl' passes termination check by lexical order 1
rev passes termination check
result: Cons(HD=C(), TL=Cons(HD=B(), TL=Cons(HD=A(), TL=Nil())))
\end{verbatim}
\runexample{nil+\%3D+Nil\%28\%29\%3B\%0D\%0Acons+\%3D+\%5Bhd\%5D\%5Btl\%5DCons\%28HD\%3Dhd\%2CTL\%3Dtl\%29\%3B\%0D\%0Al1+\%3D+cons+\%28A\%28\%29\%29+\%28cons+\%28B\%28\%29\%29+\%28cons+\%28C\%28\%29\%29+nil\%29\%29\%3B\%0D\%0A\%0D\%0Amap+\%3D+\%5Bf\%5D\%5Blist\%5Dlet\%0D\%0A\%09map\%27+\%3D+\%5Bl\%5Dcase+l+of\%0D\%0A\%09\%09\%7B+Nil+z+\%3D\%3E+Nil\%28\%29\%0D\%0A\%09\%09\%7C+Cons+pair+\%3D\%3E+Cons+\%28HD\%3D\%28f+pair.HD\%29\%2C+TL\%3D\%28map\%27+pair.TL\%29\%29\%7D\%0D\%0A\%09in+map\%27+list\%3B\%0D\%0Amap+\%28\%5Bel\%5DF\%28el\%29\%29+l1\%3B\%0D\%0A\%0D\%0Afoldl+\%3D+\%5Bf\%5D\%5Be\%5D\%5Blist\%5Dlet\%0D\%0A\%09foldl\%27+\%3D+\%5Be\%5D\%5Bl\%5Dcase+l+of\%0D\%0A\%09\%09\%7B+Nil+z+\%3D\%3E+e\%0D\%0A\%09\%09\%7C+Cons+p+\%3D\%3E+foldl\%27+\%28f+p.HD+e\%29+p.TL+\%7D\%0D\%0A\%09in+foldl\%27+e+list\%3B\%0D\%0A\%0D\%0Arev+\%3D+\%5Blist\%5Dfoldl+cons+nil+list\%3B\%0D\%0Arev+l1\%3B\%0D\%0A}

\subsection{List flattening\label{ex:flatten}}
The task is to transform a list of lists into a list, so that the
elements of the first list come first, then the elements of the second
list and so on. Example: {\tt flatten [[A,B,C],[D,E,F]] = [A,B,C,D,E,F]}.
The first version $\mathtt{flatten} : \alist(\alist \alpha)
\rightarrow \alist \alpha$ works but fails termination check
because of the limited pattern matching abilities of \foetus, but it is
also bad style and inefficient because it builds a temporary list for
the recursive call. However, the second version {\tt f} with a mutual
recursive auxiliary function $\mathtt{g} : \alist \alpha \rightarrow
\alist(\alist \alpha) \rightarrow \alist \alpha$ passes termination check.
\begin{verbatim}
nil = Nil();
cons = [hd][tl]Cons(HD=hd,TL=tl);
l1 = cons (A()) (cons (B()) (cons (C()) nil));
ll = (cons l1 (cons l1 nil));

flatten = [listlist]case listlist of
        { Nil z => Nil()
        | Cons p => case p.HD of
                { Nil z => flatten p.TL
                | Cons p' => Cons(HD=p'.HD, TL=flatten
                                (Cons(HD=p'.TL, TL=p.TL))) }};
flatten ll;

f = [l]case l of
        { Nil z => Nil()
        | Cons p => g p.HD p.TL },
g = [l][ls]case l of
        { Nil z => f ls
        | Cons p => Cons(HD=p.HD, TL=(g p.TL ls)) };
f ll;
\end{verbatim}
{\bf \foetus\ output:}\nopagebreak
\begin{verbatim}
nil passes termination check
cons passes termination check
l1 passes termination check
ll passes termination check

?: flatten -> flatten
<: flatten -> flatten
flatten FAILS termination check
result: Cons(HD=A(), TL=Cons(HD=B(), TL=Cons(HD=C(),
TL=Cons(HD=A(), TL=Cons(HD=B(), TL=Cons(HD=C(), TL=Nil()))))))

<: f -> g -> f
f passes termination check by lexical order 0
? <: g -> f -> g
< =: g -> g
g passes termination check by lexical order 1 0
result: Cons(HD=A(), TL=Cons(HD=B(), TL=Cons(HD=C(),
TL=Cons(HD=A(), TL=Cons(HD=B(), TL=Cons(HD=C(), TL=Nil()))))))
\end{verbatim}
\runexample{nil+\%3D+Nil\%28\%29\%3B\%0D\%0Acons+\%3D+\%5Bhd\%5D\%5Btl\%5DCons\%28HD\%3Dhd\%2CTL\%3Dtl\%29\%3B\%0D\%0Al1+\%3D+cons+\%28A\%28\%29\%29+\%28cons+\%28B\%28\%29\%29+\%28cons+\%28C\%28\%29\%29+nil\%29\%29\%3B\%0D\%0All+\%3D+\%28cons+l1+\%28cons+l1+nil\%29\%29\%3B\%0D\%0A\%0D\%0Aflatten+\%3D+\%5Blistlist\%5Dcase+listlist+of\%0D\%0A\%09\%7B+Nil+z+\%3D\%3E+Nil\%28\%29\%0D\%0A\%09\%7C+Cons+p+\%3D\%3E+case+p.HD+of\%0D\%0A\%09\%09\%7B+Nil+z+\%3D\%3E+flatten+p.TL\%0D\%0A\%09\%09\%7C+Cons+p\%27+\%3D\%3E+Cons\%28HD\%3Dp\%27.HD\%2C+TL\%3Dflatten+\%28Cons\%28HD\%3Dp\%27.TL\%2C+\%0D\%0A\%09\%09\%09\%09\%09\%09\%09TL\%3Dp.TL\%29\%29\%29+\%7D\%7D\%3B\%0D\%0Aflatten+ll\%3B\%0D\%0A\%0D\%0Af+\%3D+\%5Bl\%5Dcase+l+of\%0D\%0A\%09\%7B+Nil+z+\%3D\%3E+Nil\%28\%29+\%0D\%0A\%09\%7C+Cons+p+\%3D\%3E+g+p.HD+p.TL+\%7D\%2C\%0D\%0Ag+\%3D+\%5Bl\%5D\%5Bls\%5Dcase+l+of\%0D\%0A\%09\%7B+Nil+z+\%3D\%3E+f+ls\%0D\%0A\%09\%7C+Cons+p+\%3D\%3E+Cons\%28HD\%3Dp.HD\%2C+TL\%3D\%28g+p.TL+ls\%29\%29+\%7D\%3B\%0D\%0Af+ll\%3B\%0D\%0A}

\subsection{Merge sort}
With type
$$
    \bool := \{ \mathtt{True()} | \mathtt{False()} \}
$$
we can define {\tt le\_nat}$: \nat \rightarrow \nat \rightarrow
\bool$ and
$\mathtt{merge} : (\alpha \rightarrow \alpha \rightarrow
\bool) \rightarrow \alist \alpha \rightarrow \alist \alpha \rightarrow
\alist \alpha$ as follows:
\begin{verbatim}
merge = [le][l1][l2]case l1 of
        { Nil z => l2
        | Cons p1 => case l2 of
                { Nil z => l1
                | Cons p2 => case (le p1.HD p2.HD) of
                        { True  z => Cons(HD=p1.HD,
                                     TL=merge le p1.TL l2)
                        | False z => Cons(HD=p2.HD,
                                     TL=merge le l1 p2.TL) }}};

le_nat = [x][y]case x of
        { O z  => True()
        | S x' => case y of
                { O z  => False()
                | S y' => le_nat x' y' }};

i = S(O());
ii = S(S(O()));
iii = S(S(S(O())));
iv = S(S(S(S(O()))));
v = S(S(S(S(S(O())))));
l1 = Cons(HD=O(), TL=Cons(HD=iii, TL=Cons(HD=iv, TL=Nil())));
l2 = Cons(HD=i,   TL=Cons(HD=ii,  TL=Cons(HD=v,  TL=Nil())));
merge le_nat l1 l2;
\end{verbatim}
{\bf \foetus\ output:}\nopagebreak
\begin{verbatim}
= < <: merge -> merge -> merge
= = <: merge -> merge
= < =: merge -> merge
merge passes termination check by lexical order 1 2
< <: le_nat -> le_nat
le_nat passes termination check by lexical order 0
result: Cons(HD=O(), TL=Cons(HD=S(O()), TL=Cons(HD=S(S(O())),
     TL=Cons(HD=S(S(S(O()))), TL=Cons(HD=S(S(S(S(O())))),
     TL=Cons(HD=S(S(S(S(S(O()))))), TL=Nil()))))))
\end{verbatim}
\runexample{merge+\%3D+\%5Ble\%5D\%5Bl1\%5D\%5Bl2\%5Dcase+l1+of\%0D\%0A\%09\%7B+Nil+z+\%3D\%3E+l2\%0D\%0A\%09\%7C+Cons+p1+\%3D\%3E+case+l2+of\%0D\%0A\%09\%09\%7B+Nil+z+\%3D\%3E+l1\%0D\%0A\%09\%09\%7C+Cons+p2+\%3D\%3E+case+\%28le+p1.HD+p2.HD\%29+of\%0D\%0A\%09\%09\%09\%7B+True++z+\%3D\%3E+Cons\%28HD\%3Dp1.HD\%2C+TL\%3Dmerge+le+p1.TL+l2\%29\%0D\%0A\%09\%09\%09\%7C+False+z+\%3D\%3E+Cons\%28HD\%3Dp2.HD\%2C+TL\%3Dmerge+le+l1+p2.TL\%29+\%7D\%7D\%7D\%3B\%0D\%0A\%0D\%0Ale_nat+\%3D+\%5Bx\%5D\%5By\%5Dcase+x+of\%0D\%0A\%09\%7B+O+z++\%3D\%3E+True\%28\%29\%0D\%0A\%09\%7C+S+x\%27+\%3D\%3E+case+y+of\%0D\%0A\%09\%09\%7B+O+z++\%3D\%3E+False\%28\%29\%0D\%0A\%09\%09\%7C+S+y\%27+\%3D\%3E+le_nat+x\%27+y\%27+\%7D\%7D\%3B\%0D\%0A\%0D\%0Ai+\%3D+S\%28O\%28\%29\%29\%3B\%0D\%0Aii+\%3D+S\%28S\%28O\%28\%29\%29\%29\%3B\%0D\%0Aiii+\%3D+S\%28S\%28S\%28O\%28\%29\%29\%29\%29\%3B\%0D\%0Aiv+\%3D+S\%28S\%28S\%28S\%28O\%28\%29\%29\%29\%29\%29\%3B\%0D\%0Av+\%3D+S\%28S\%28S\%28S\%28S\%28O\%28\%29\%29\%29\%29\%29\%29\%3B\%0D\%0Al1+\%3D+Cons\%28HD\%3DO\%28\%29\%2C+TL\%3DCons\%28HD\%3Diii\%2C+TL\%3DCons\%28HD\%3Div\%2C+TL\%3DNil\%28\%29\%29\%29\%29\%3B\%0D\%0Al2+\%3D+Cons\%28HD\%3Di\%2C+++TL\%3DCons\%28HD\%3Dii\%2C++TL\%3DCons\%28HD\%3Dv\%2C++TL\%3DNil\%28\%29\%29\%29\%29\%3B\%0D\%0Amerge+le_nat+l1+l2\%3B\%0D\%0A}

\subsection{Parameter permutation: list zipping\label{ex:parmperm}}
The following function $\mathtt{zip} : \alist \alpha \rightarrow
\alist \alpha \rightarrow \alist \alpha$ combines two lists into one by
alternately taking the first elements form these lists and putting
them into the result list.
\begin{verbatim}
zip = [l1][l2]case l1 of
        { Nil z => l2
        | Cons p1 => Cons(HD=p1.HD, TL=zip l2 p1.TL) };

zip (Cons(HD=A(), TL=Cons(HD=C(), TL=Nil())))
    (Cons(HD=B(), TL=Cons(HD=D(), TL=Nil())));
\end{verbatim}
{\bf \foetus\ output:}\nopagebreak
\begin{verbatim}
? ?: zip -> zip -> zip -> zip
< <: zip -> zip -> zip
? ?: zip -> zip
zip FAILS termination check
result: Cons(HD=A(), TL=Cons(HD=B(), TL=Cons(HD=C(),
     TL=Cons(HD=D(), TL=Nil()))))
\end{verbatim}
\runexample{zip+\%3D+\%5Bl1\%5D\%5Bl2\%5Dcase+l1+of\%0D\%0A\%09\%7B+Nil+z+\%3D\%3E+l2\%0D\%0A\%09\%7C+Cons+p1+\%3D\%3E+Cons\%28HD\%3Dp1.HD\%2C+TL\%3Dzip+l2+p1.TL\%29+\%7D\%3B\%0D\%0A+\%0D\%0Azip+\%28Cons\%28HD\%3DA\%28\%29\%2C+TL\%3DCons\%28HD\%3DC\%28\%29\%2C+TL\%3DNil\%28\%29\%29\%29\%29\%0D\%0A++++\%28Cons\%28HD\%3DB\%28\%29\%2C+TL\%3DCons\%28HD\%3DD\%28\%29\%2C+TL\%3DNil\%28\%29\%29\%29\%29\%3B\%0D\%0A}
Here in the recursion of {\tt zip} one arguments is decreasing, but
arguments are switched. Thus only a even number of recursive calls
produces a structural decrease on {\tt l1} and {\tt l2}.
\foetus\ does not recognize {\tt zip} to be terminating because not
{\em every} (direct or indirect) recursive call makes the arguments smaller
on any structural lexical order.

Of course there are simple orders
that fulfill the demanded criteria, like $<$ on {$|l1|+|l2|$}.
Another solution is to ``copy'' {\tt zip} into {\tt zip'} and implement
{\em mutual recursion} as follows:
\begin{verbatim}
zip = [l1][l2]case l1 of
        { Nil z => l2
        | Cons p1 => Cons(HD=p1.HD, TL=zip' l2 p1.TL) },
zip'= [l1][l2]case l1 of
        { Nil z => l2
        | Cons p1 => Cons(HD=p1.HD, TL=zip l2 p1.TL) };

zip (Cons(HD=A(), TL=Cons(HD=C(), TL=Nil())))
    (Cons(HD=B(), TL=Cons(HD=D(), TL=Nil())));
\end{verbatim}
{\bf \foetus\ output:}\nopagebreak
\begin{verbatim}
< <: zip -> zip' -> zip
zip passes termination check by lexical order 0
< <: zip' -> zip -> zip'
zip' passes termination check by lexical order 0
result: Cons(HD=A(), TL=Cons(HD=B(), TL=Cons(HD=C(), TL=Cons(HD=D(),
TL=Nil()))))
\end{verbatim}
\runexample{zip+\%3D+\%5Bl1\%5D\%5Bl2\%5Dcase+l1+of\%0D\%0A\%09\%7B+Nil+z+\%3D\%3E+l2\%0D\%0A\%09\%7C+Cons+p1+\%3D\%3E+Cons\%28HD\%3Dp1.HD\%2C+TL\%3Dzip\%27+l2+p1.TL\%29+\%7D\%2C\%0D\%0Azip\%27\%3D+\%5Bl1\%5D\%5Bl2\%5Dcase+l1+of\%0D\%0A\%09\%7B+Nil+z+\%3D\%3E+l2\%0D\%0A\%09\%7C+Cons+p1+\%3D\%3E+Cons\%28HD\%3Dp1.HD\%2C+TL\%3Dzip+l2+p1.TL\%29+\%7D\%3B\%0D\%0A+\%0D\%0Azip+\%28Cons\%28HD\%3DA\%28\%29\%2C+TL\%3DCons\%28HD\%3DC\%28\%29\%2C+TL\%3DNil\%28\%29\%29\%29\%29\%0D\%0A++++\%28Cons\%28HD\%3DB\%28\%29\%2C+TL\%3DCons\%28HD\%3DD\%28\%29\%2C+TL\%3DNil\%28\%29\%29\%29\%29\%3B\%0D\%0A}

\subsection{Tuple parameter\label{ex:tupelparm}}
This example, an alternative version of $\mathtt{add} :
\mathtt{(X} : \nat \mathtt{, Y} : \nat \mathtt{)} \rightarrow \nat$, shows that \foetus\
loses dependency information if you ``pack'' and ``unpack'' tuples.
\begin{verbatim}
add = [xy]case xy.X of
        { O z => xy.Y
        | S x' => S(add (X=x', Y=xy.Y)) };
\end{verbatim}
{\bf \foetus\ output:}\nopagebreak
\begin{verbatim}
?: add -> add
add FAILS termination check
\end{verbatim}
\runexample{add+\%3D+\%5Bxy\%5Dcase+xy.X+of+\%0D\%0A\%09\%7B+O+z+\%3D\%3E+xy.Y\%0D\%0A\%09\%7C+S+x\%27+\%3D\%3E+S\%28add+\%28X\%3Dx\%27\%2C+Y\%3Dxy.Y\%29\%29+\%7D\%3B\%0D\%0A}

\subsection{Transfinite addition of ordinal numbers\label{ex:addord}}
The type of ordinal numbers is
$$
   \ord := \Let{\ord = \{ \mathtt{O()} | \mathtt{S(} \ord \mathtt{)} |
        \mathtt{Lim(} \nat \rightarrow \ord \mathtt{)} \} }{\ord}
$$
and $\mathtt{addord} : \ord \rightarrow \ord \rightarrow \ord$ can be
implemented as follows:
\begin{verbatim}
addord = [x][y]case x of
        { O o => y
        | S x' => S(addord x' y)
        | Lim f => Lim([z]addord (f z) y) };
\end{verbatim}
{\bf \foetus\ output:}\nopagebreak
\begin{verbatim}
< =: addord -> addord
addord passes termination check by lexical order 0
\end{verbatim}
\runexample{addord+\%3D+\%5Bx\%5D\%5By\%5Dcase+x+of+\%0D\%0A\%09\%7B+O+o+\%3D\%3E+y\%0D\%0A\%09\%7C+S+x\%27+\%3D\%3E+S\%28addord+x\%27+y\%29+\%0D\%0A\%09\%7C+Lim+f+\%3D\%3E+Lim\%28\%5Bz\%5Daddord+\%28f+z\%29+y\%29+\%7D\%3B\%0D\%0A}

\subsection{Fibonacci numbers\label{ex:fib}}
Iterative version
$\mathtt{fib}: \nat \rightarrow \nat$
of algorithm to calculate the fibonacci numbers
$\mathrm{fib}(0)=1$, $\mathrm{fib}(1)=1$, 2, 3, 5, 8, $\dots$. Only
the first parameter is important for termination, the
second and the third parameter are ``accumulators''.
\begin{verbatim}
fib' = [n][fn][fn']case n of
        { O z  => fn
        | S n' => fib' n' (add fn fn') fn};
fib = [n]fib' n (S(O())) (O());
\end{verbatim}
{\bf \foetus\ output:}\nopagebreak
\begin{verbatim}
< ? ?: fib' -> fib' -> fib'
< ? ?: fib' -> fib'
fib' passes termination check by lexical order 0
fib passes termination check
\end{verbatim}
\runexample{add+\%3D+\%5Bx\%5D\%5By\%5Dcase+x+of+\%7B+O+z+\%3D\%3E+y+\%7C+S+x\%27+\%3D\%3E+S\%28add+x\%27+y\%29+\%7D\%3B\%0D\%0A\%0D\%0Afib\%27+\%3D+\%5Bn\%5D\%5Bfn\%5D\%5Bfn\%27\%5Dcase+n+of\%0D\%0A+\%7B+O+z+\%3D\%3E+fn\%0D\%0A+\%7C+S+n\%27+\%3D\%3E+fib\%27+n\%27+\%28add+fn+fn\%27\%29+fn\%7D\%3B\%0D\%0Afib+\%3D+\%5Bn\%5Dfib\%27+n+\%28S\%28O\%28\%29\%29\%29+\%28O\%28\%29\%29\%3B\%0D\%0A\%0D\%0A\%28fib+\%28S\%28S\%28S\%28S\%28S\%28O\%28\%29\%29\%29\%29\%29\%29\%29\%29\%3B+\%0D\%0A}

\subsection{Non-terminating mutual recursion}
The following three functions $\mathtt{f}, \mathtt{g}, \mathtt{h} :
\nat \rightarrow \nat \rightarrow \nat$ are
an artificial example for non-termination that has been designed to
show to what extent the call graph has to be completed to assure
correct results of the termination checker. Function {\tt h}
(here $h(x,y)=0 \;\forall
x, y$) could be any function
that ``looks into'' its arguments, e.g. {\tt add}.
\begin{verbatim}
h = [x][y]case x of
        { O z  => case y of
                { O z  => O()
                | S y' => h x y' }
        | S x' => h x' y },

f = [x][y]case x of
        { O z  => O()
        | S x' => case y of
                { O z  => O()
                | S y' => h (g x' y) (f (S(S(x))) y') } },

g = [x][y]case x of
        { O z  => O()
        | S x' => case y of
                { O z  => O()
                | S y' => h (f x y) (g x' (S(y))) } };

(* f (S(S(O()))) (S(S(O()))); *)
\end{verbatim}
{\bf \foetus\ output:} Note that the combined call $f \rightarrow g
\rightarrow f$ still does not prevent termination. But then call graph
completion finds $f \rightarrow g \rightarrow g \rightarrow f$ that
destroys the lexical order 1 0 that was possible until then.
\begin{verbatim}
< <: h -> h -> h
< =: h -> h
= <: h -> h
h passes termination check by lexical order 0 1
< ?: f -> g -> g -> f
? ?: f -> f -> g -> g -> f
< =: f -> g -> f
? <: f -> f
f FAILS termination check
? <: g -> f -> f -> g
? ?: g -> g -> f -> f -> g
< =: g -> f -> g
< ?: g -> g
g FAILS termination check
\end{verbatim}
\runexample{h+\%3D+\%5Bx\%5D\%5By\%5Dcase+x+of\%0D\%0A\%09\%7B+O+z++\%3D\%3E+case+y+of\%0D\%0A\%09\%09\%7B+O+z++\%3D\%3E+O\%28\%29\%0D\%0A\%09\%09\%7C+S+y\%27+\%3D\%3E+h+x+y\%27+\%7D\%0D\%0A\%09\%7C+S+x\%27+\%3D\%3E+h+x\%27+y+\%7D\%2C\%0D\%0A\%0D\%0Af+\%3D+\%5Bx\%5D\%5By\%5Dcase+x+of+\%0D\%0A\%09\%7B+O+z++\%3D\%3E+O\%28\%29\%0D\%0A\%09\%7C+S+x\%27+\%3D\%3E+case+y+of+\%0D\%0A\%09\%09\%7B+O+z++\%3D\%3E+O\%28\%29\%0D\%0A\%09\%09\%7C+S+y\%27+\%3D\%3E+h+\%28g+x\%27+y\%29+\%28f+\%28S\%28S\%28x\%29\%29\%29+y\%27\%29+\%7D+\%7D\%2C\%0D\%0A\%0D\%0Ag+\%3D+\%5Bx\%5D\%5By\%5Dcase+x+of+\%0D\%0A\%09\%7B+O+z++\%3D\%3E+O\%28\%29+\%0D\%0A\%09\%7C+S+x\%27+\%3D\%3E+case+y+of+\%0D\%0A\%09\%09\%7B+O+z++\%3D\%3E+O\%28\%29\%0D\%0A\%09\%09\%7C+S+y\%27+\%3D\%3E+h+\%28f+x+y\%29+\%28g+x\%27+\%28S\%28y\%29\%29\%29+\%7D+\%7D\%3B\%0D\%0A\%0D\%0A\%28*+f+\%28S\%28S\%28O\%28\%29\%29\%29\%29+\%28S\%28S\%28O\%28\%29\%29\%29\%29\%3B+*\%29\%0D\%0A}





\section{Termination Checker Overall Outline\label{sec:outline}}
%

\subsection{Function call extraction\label{sec:funex}}

The task of \foetus\ is to check whether functions terminate or
not. Because the \foetus\ language is functional and no direct loop
constructs  exist, the only means to form loops is
recursion. Therefore out of the program text all function calls have
to be extracted to find direct or indirect recursive calls that may
cause termination problems.

The heart of \foetus\ is a analyzer that runs through the syntax tree of
the given \foetus\ program and looks for {\em applications}.
Consecutive applications are gathered and formed in to a
{\em function call}, e.g. in example \ref{ex:add}, function {\tt add}.
There the two applications \ttverb{((add x') y)} form the call
$\mathrm{add}(x',y)$. As you see in this example ``add'' is always terminating
because in each recursive call the first argument $x$ is
decreased. \foetus\ stores with each call information about how the
arguments of the call ($x'$, $y$ in the example) relate to the parameters
of the calling function (here: $x$, $y$), the so-called depedencies (here:
$x'<x$, $y=y$). We distinguish three kinds of relations: $\less$
(less), $\equal$ (equal) and $\unknown$ (unknown, this includes `greater').

The abilities of \foetus\ to recognise dependencies are yet very
limited. So far only three cases are considered:
\begin{enumerate}
\item Constructor elimination.\label{rule:CEl}\\
Be $x$, $y$ variables and $C$ a constructor, and $x=C(y)$. It follows
$y \less x$. This is applied in case constructs (see example above).
\item Projection.\label{rule:Proj}\\
Be $x$, $y$ variables, $L$ a label, $\rho$ a relation in $\{\less,
\equal\}$ and $y \,\rho\, x$. Here it follows $y.L \,\rho\, x$, i.d. a
component is considered as big as the entire tuple.
\item Application.\label{rule:App}\\
Be $x$, $y$ variables, $a$ a vector of terms (arguments of $y$),
$\rho$ a relation in $\{\less, \equal\}$ and $y \,\rho\, x$. It
follows $(y a) \,\rho\, x$.
\end{enumerate}
The rule \ref{rule:App} may have a strange looking, but it can be
applied in example \ref{ex:addord} ({\tt addord}).
In the third case $x = \mathrm{Lim}(f)$ we have with rule
\ref{rule:CEl} $f \infixless x$ and with rule \ref{rule:App}
$(f z) \infixless x$, therefore {\tt addord} is terminating.

\subsection{Call graph}
In the end the whole of extracted function calls form the {\em call
  graph}. It is a multigraph; each vertex represents a function and each
edge from vertex $f$ to vertex $g$ a call of function $g$ within the
function of $f$. The edges are labeled with the dependency information
(see above) put in a {\em call matrix}. The call matrix for the only
one call $\mathrm{add}\rightarrow\mathrm{add}$ in example \ref{ex:add}
would be
\begin{center}
\begin{tabular}{c|cc}
     &   $x$      &    $y$ \\
\hline
$x'$ & $\less$    & $\unknown$ \\
$y$  & $\unknown$ & $\equal$
\end{tabular}
\end{center}
Note that each row represents one call argument and its relations to
the calling function parameters.

Now if a function $f$ calls a function $g$ and the latter calls
another function $h$, $f$ indirectly calls $h$. The call matrix of
this {\em combined call} $f \rightarrow h$ is the product of the two
matrices of $g \rightarrow h$ and  $f \rightarrow g$. We get the
{\em completed call graph} if we insert all combined calls (as new
edges) into the original graph.

To find out whether a function $f$ is terminating you have to collect
all calls from $f$ to itself out of the completed call graph (this
includes the direct and the indirect calls). When a {\em lexical order}
exists on the function parameters of $f$ so that every recursive call
decreases the order of the parameters, we have proven the termination
of f. This order we call {\em termination order}.

We could call the algorithms of call graph completion and finding a
lexical order the ``brain'' of \foetus; it is described more precisely
and formally in the next section.


\section{Formal Description\label{sec:formal}}

\subsection{Call Matrix}
Be $R = \{\less, \equal, \unknown\}$ set of the relations ``less
than'',  ``equal to'' and ``relation unknown''. In the context of
``$f(x,y)$ calls $g(a,b)$'' $a \infixless y$ means ``we know that (call)
argument $a$ is less than (input) parameter $y$'', $a \infixequal y$ means
``$a$ is (at least) equal to $y$ (if not less than)'' and $a \infixunknown
y$ means ``we do not know the relation between $a$ and $y$''.

With the two operations
$+$ and $\cdot$ defined as in table \ref{tabOpR} $R$ forms a
commutative
rig\footnote{On the WWW I found the English term
    ``rig'' for what Germans call a ``Halb\-ring''. This is probably a
    play of words: Compared  to a ``ring'' a ``rig'' misses an ``n''
    as well as inverse elements regarding addition. I cite Ross Moore
(see {\tt \htmladdnormallink{http://www.mpce.mq.edu.au/\tttilde{}ross/maths/Quantum/Sect1.html\#206}{http://www.mpce.mq.edu.au/~ross/maths/Quantum/Sect1.html}}):
\begin{quote}
A {\em rig} is a set $R$ enriched with two monoid structures, a
commutative one written additively and the other written
multiplicatively, such that the following equations hold:
$$
            a 0 = 0 = 0 a
$$
$$
    a (b + c) = a b + a c, \qquad (a + b) c = a c + a b
$$
The natural numbers $\N$ provide an example of a rig.

A {\em ring} is a rig for which the additive monoid is a group. The
integers  $\Z$ provide an example.

A rig is {\em commutative}  when the multiplicative monoid is commutative.
\end{quote}
}
with $0$-element $\unknown$ and $1$-element $\equal$. The operation $+$ can be
understood as ``combining {\em parallel} information about a relation'',
e.g. if we have $a \infixunknown y$ and $a \infixless y$ we have $a
\,(\unknown + \less)\, y$ and that simplifies to $a \infixless y$.
The operation
$\cdot$ however is ``{\em serial} combination'', e.g. $a \infixless y$
and $y \infixequal z$ can be combined into $a \,(\less \cdot \equal)\, z$,
simplified: $a \infixless z$. $\unknown$ is neutral regarding $+$ because
it gives you no new information, whereas $\less$ is dominant because
it is the strongest information. Regarding $\cdot$ the relation
$\equal$ is neutral
and $\unknown$ is dominant because it ``destroys'' all
information. Check the table to see which relation overrides which.

\begin{table}[h]
  \begin{center}
    \leavevmode
\hbox{
\hbox{
\begin{tabular}{c|ccc}
$+$        & $\less$ & $\equal$ & $\unknown$ \\
\hline
$\less$    & $\less$ & $\less$  & $\less$ \\
$\equal$   & $\less$ & $\equal$ & $\equal$ \\
$\unknown$ & $\less$ & $\equal$ & $\unknown$
\end{tabular}
}%
\hspace{2cm}%
\hbox{
\begin{tabular}{c|ccc}
$\cdot$    & $\less$    & $\equal$   & $\unknown$ \\
\hline
$\less$    & $\less$    & $\less$    & $\unknown$ \\
$\equal$   & $\less$    & $\equal$   & $\unknown$ \\
$\unknown$ & $\unknown$ & $\unknown$ & $\unknown$
\end{tabular}%
}
}
    \caption{Operations on $R$}
    \label{tabOpR}
  \end{center}
\end{table}

Now we can define multiplication on matrices over R as usual:
$$
    \cdot : R^{n \times m} \times R^{m \times l} \rightarrow R^{n
      \times l}
$$
$$
    \left((a_{ij}), (b_{ij})\right) \mapsto (c_{ij}) = \left(
      \sum_{k=1}^m a_{ik} b_{kj} \right)
$$
Why is this a reasonable definition? Assume you have three sets of
variables $\{x_1, \dots, x_n\}$, $\{y_1, \dots, y_m\}$ and $\{z_1,
\dots, z_l\}$, a matrix $A=(a_{ij}) \el R^{n \times m}$ reflecting the
relations between the $x_i$s and the $y_i$s (i.d. $a_{ij} = \rho \iff
x_i \,\rho\, y_j$) and a matrix $B \el R^{m \times l}$ reflecting the
relations between the $y_i$s and the $z_i$s. Then the matrix product
$C=AB$ reflects the relations between the $x_i$s and the $z_i$s. Because
$$
    c_{ij} = a_{i1} \cdot b_{1j} + a_{i2} \cdot b_{2j} + \dots +
    a_{im} \cdot b_{mj},
$$
we have e.g. $x_i \infixless z_j$ if we know it by intermediate variable $y_1$
($a_{i1} \cdot b_{1j} = \quotless$) {\em or} by intermediate variable
$y_2$ {\em or} $\dots$ (to be continued).

\paragraph*{Definition.}
A {\em call matrix} is a matrix over $R$ with no more than one element
different from $\unknown$ per row.
$$
    \mathrm{CM}(n,m) := \{ (a_{ij}) \el R^{n \times m} : \forall i
    \forall j \forall k \not= j ( a_{ij} = \unknown \lor a_{ik} = \unknown) \}
$$
\paragraph*{Remark.} The reason we define call matrices this way is
these are the only ones \foetus\ produces by function call extraction
(see section \ref{sec:funex}). Because \foetus\ recognizes only the
three described cases of dependecies, a call argument can only depend
of {\em one} function parameter. But multiple dependecies are
imaginable, like in
\begin{verbatim}
f(x,y) = if (x=0) then 0 else let a=min(x,y)-1 in f(a,x)
\end{verbatim}
Here the second call argument {\tt a} is less than both {\tt x} and
{\tt y}. The next proposition assures that all matrices \foetus\ will
have do deal with are {\em call} matrices.
\paragraph*{Proposition.}
Matrix multiplication on matrices induces a multiplication on call matrices
$$
    \cdot : \mathrm{CM}(n,m) \times \mathrm{CM}(m,l) \rightarrow
    \mathrm{CM}(n,l)
$$
This operation is well defined.
\paragraph*{Proof.}
Be $A = (a_{ij}) \el \mathrm{CM}(n,m)$, $B = (b_{ij}) \el
\mathrm{CM}(m,l)$, $AB = C = (c_{ij}) \el R^{n \times l}$ and $k(i)$ the
index of the element of the $i$th row of $A$ that is different to
$\unknown$ (or 1, if no such element exists). The we have with the
rules in rig $R$
$$
    c_{ij} = \sum_{k=1}^m a_{ik} b_{kj} = a_{i,k(i)} b_{k(i),j}
$$
Now consider the $i$th row of $C$:
$$
    c_i = (c_{ij})_{1 \leq j \leq l} = (a_{i,k(i)} b_{k(i),j})_{1 \leq
      j \leq l}
$$
Because at most one $b_{k(i),j}$ is unequal to $\unknown$, at most one
element of $c_i$ is unequal to $\unknown$. Therefore $C \el
\mathrm{CM}(n,l)$.

\subsection{Call Graph}
For each $i \el \N$ we assume a set $F^{(i)} = \{f^{(i)}, g^{(i)},
h^{(i)}, ...\}$ of identifiers for functions of arity $i$,
$\mathcal{F} = \biguplus_{i \el \N} F^{(i)}$.
\paragraph*{Definition.}
We form the set of {\em calls} as follows
$$
   \mathcal{C} = \{ (f^{(n)}, g^{(m)}, A) : f^{(n)} \el F^{(n)},
   g^{(m)} \el F^{(m)}, A \el \mathrm{CM}(m,n) \}
$$
On calls we define the partial operation {\em combination of calls}
$$
   \circ : \mathcal{C} \times \mathcal{C} \rightarrow \mathcal{C}
$$
$$
   \left( (g^{(m)}, h^{(l)}, B), (f^{(n)}, g^{(m)}, A) \right) \mapsto
   (f^{(n)}, h^{(l)}, BA)
$$
Meaning: If $g$ calls $h$ with call matrix $B$ and $f$ calls $g$ with
call matrix $A$, then $f$ indirectly calls $h$ with call matrix
$BA$. $\circ$ cannot be applied to calls that have no ``common
function'' like $g$, therefore it is partial. $\circ$ can be expanded
to sets of calls
$$
   \circ : \mathcal{P}(\mathcal{C}) \times \mathcal{P}(\mathcal{C})
   \rightarrow \mathcal{P}(\mathcal{C})
$$
$$
   (C, C') \mapsto \{ c \circ_{\mathcal{C}} c' : c \el C,
       c' \el C',
       (c,c') \el \mathrm{Dom}(\circ_{\mathcal{C}}) \}
$$
Here we combine each call in $C$ with each call in $\hat C$ to which
$\circ_{\mathcal{C}}$ is applicable and form a set of the combined
calls. $\circ_{\mathcal{P}(\mathcal{C})}$ is a total function.
\paragraph*{Definition.}
A {\em call graph} is a graph $(V,E)$ with
vertices $V = \mathcal F$ and edges $E \finitsubset \mathcal{C}$.
A call graph is {\em complete} if
$$
    E \circ E \subseteq E
$$

\paragraph*{Definition.}
The {\em completion} of a call graph $(V,E)$ is a call graph $(V,E')$
such that
\renewcommand{\labelenumi}{(\arabic{enumi})}
\begin{enumerate}
\item $(V,E')$ is complete,\label{complete1}
\item $E \subseteq E'$ and\label{complete2}
\item for all $E''$ satisfying (\ref{complete1}) and
  (\ref{complete2}) we have $E' \subseteq E''$.
\end{enumerate}

\paragraph*{Proposition.}
The completion of a call graph $(V,E)$ is the call graph $(V,E')$
such that
$$
    c \el E' \iff \exists n>0, c_1, \dots, c_n \el E : c_1 \circ \dots \circ
    c_n = c
$$
\paragraph*{Proof.}
\begin{enumerate}
\item
Be $c \el E' \circ E'$. Then there are $d,e \el E'$ with $c = d \circ
e$. Because $(V,E')$ is complete, we have
\begin{eqnarray*}
    d = d_1 \circ \dots \circ d_n & \qquad & d_1,\dots,d_n \el E \\
    e = e_1 \circ \dots \circ e_m & \qquad & e_1,\dots,e_m \el E
\end{eqnarray*}
Thus $c = d_1 \circ \dots \circ d_n \circ e_1 \circ \dots \circ e_m
\el E'$.
\item $E \subseteq E'$ is trivial with $n=1$.
\item Be $(V,E'')$ complete and $E \subseteq E''$. This gives us $E \circ
  E \subseteq E''$ from which we gain by induction
$$
    \underbrace{E \circ \dots \circ E}_{n\mathrm{-times}} =:
    E^n \subseteq E'' \;\;\mathrm{for}\;\mathrm{all}\;n.
$$
Now be $c \el E'$. That implies $c = c_1 \circ \dots \circ c_n$ ($c_i
\el E$) for a suitable $n$. Hence $c \el E^n \subseteq E''$.
\qed
\end{enumerate}


\paragraph*{Proposition. (Completion algorithm) }
Be $(V,E)$ a call graph, $(V,E')$ its completion and $(E_n)_{n \el
  \N}$  a sequence of sets of calls defined as follows:
\begin{eqnarray*}
    E_0     & = & E \\
    E_{n+1} & = & E_n \cup (E_n \circ E)
\end{eqnarray*}
Then there is a $n \el \N$ so that
$$
    E' = E_n = E_{n+1} = E_{n+2} = \dots
$$
(Obviously the $E_n$ grow monotonously.)
\paragraph*{Proof.}
First we show by induction that $E_n \subseteq E'$ for all $n \el
\N$: It is obvious that $E_0 \subseteq E'$. Now be $E_n \subseteq E'$ and
$c \el E_{n+1} \setminus E_n$. Then $c \el E_n \circ E$, therefore $c
= d_1 \circ \dots \circ d_n \circ e$, $d_1, \dots, d_n, e \el E$. It
follows $c \el E'$, $E_{n+1} \subseteq E'$.

Second: Because we have a finit set of starting edges $E$ and
therefore a finit set of reachable vertices and also a finit set of
possible edges between two vertices (limited by the number of
different call matrices of fixed dimensions) the $E_i$s cannot grow
endlessly. Thus an $n \el \N$ exists with $E_n = E_{n+1}$.

Third: We show that $E' \subseteq E_n$ for that particular $n$. Be $c
\el E'$. Then there exists an $m$ such that $c = d_1 \circ \dots \circ
d_m$, therefore $c \el E_m$. Now if $m \leq n$ then $E_m \subseteq
E_n$, otherwise $m > n$ and hence $E_m = E_n$, in both cases $c \el E_n$.
\qed

\subsection{Lexical Order}
\paragraph*{Definition.} Be $(V,E)$ a complete call graph and
$f^{(i)}$ a function of arity $i$. We call
$$
    E_{f^{(i)}} := \{ \Delta(C): (f^{(i)}, f^{(i)}, C) \el E \}
    \subset R^i
$$
the {\em recursion behaviour} of function $f^{(i)}$. ($\Delta$ takes the
diagonal of square matrices).

Each row of this set represents one possible recursive call of $f^{(i)}$
and how the orders of all parameters are altered in this call. The
diagonals of the call matrices are taken because we want to know only
how a parameter relates to its old value in the last call to
$f^{(i)}$. $E_{f^{(i)}}$ of course is a {\em finite} subset of $R^i$.

In the following we identify lexical orders on parameters with
permutations $\pi \el S_n$ of the arguments. Often not all of the
parameters are relevant for termination; these are not listed in the
lexical order and can appear in the permutation in any sequence.

In example \ref{ex:fib} ({\tt fib'}) only the argument 0 has to be
considered to prove termination, the order of argument 1 and 2 are
irrelevant and therefore both permutations
$$
    \pi_1 = \left(\begin{array}{ccc}0&1&2\\ 0&1&2\end{array}\right)
$$
and
$$
    \pi_2 = \left(\begin{array}{ccc}0&1&2\\ 0&2&1\end{array}\right)
$$
are valid continuations of the lexical order ``0''.
\paragraph*{Note:}
In the following we abbreviate the notation of permutations to
$\pi_1 = [0 1 2]$ and $\pi_2 = [0 2 1]$.

\paragraph*{Definition. (1)}
Be $B$ the recursion behaviour of function $f^{(n)}$.
We call the permutation $\pi \el S_n$ a {\em termination order} for
$f^{(n)}$ if
$$
    \forall r \el B \exists 1 \leq k \leq n
    : r_{\pi(k)} = \quotless \land (\forall 1 \leq i \leq k :
    r_{\pi(i)} = \quotequal)
$$

This definition is a very wide one. In most cases you will look for
more special termination orders:

\paragraph*{Definition. (2, inductive)}
Be $B$ the recursion behaviour of a given function.
We call the permutation $\pi \el S_n$ a {\em termination order} on $B$
if $|B| = 0$ or
\begin{eqnarray}
    && \exists r \el B : r_{\pi(0)} = \less \nonumber\\
    && \land \not\exists r \el B : r_{\pi(0)} = \unknown
    \nonumber\\
    &&  \land \,\pi'_{0} \el S_{n-1} \;\mathrm{termination}\;\mathrm{order}
    \;\mathrm{on}\; B' := \{r'_{\pi(0)} : r_{\pi(0)} \not= \less\} \subset
    R^{n-1} \nonumber
\end{eqnarray}
whereas $\pi'_i = [k_0 \dots k_{i-1} k_{i+1} \dots k_{n-1}] \el S_{n-1}$ given
$\pi = [k_0 \dots k_{n-1}] \el S_n$ and $r'_i = (k_0, \dots, k_{i-1},
k_{i+1}, \dots, k_{n-1}) \el R^{n-1}$ given $r = (k_0, \dots, k_{n-1})
\el R^n$.


The algorithm implemented in \foetus\ searches termination orders
like in definition (2); it is a one-to-one transfer of this
definition. Every termination order matching definition (2) also
matches definition (1) and it can easily be shown that if there is a
termination order of type (1) there also exists on of type (2).
\begin{example}
Be $E = \{(\equal, \less, \unknown), (\equal, \equal, \less), (\equal,
\less, \equal)\}$ the given recursion behaviour. Then $\pi_1 = [0 1
2]$ is a type (1) termination order on $E$ and $\pi_2 = [1 2 0]$ is of
both types.
\end{example}


\section{Implementation\label{sec:implementation}}
\foetus\ has been implemented in
\htmladdnormallink{SML
  97}{http://cm.bell-labs.com/cm/cs/what/smlnj/sml97.html}. We have used the
new \htmladdnormallink{Standard ML Basis
  Library}{http://www.dina.kvl.dk/\%7Esestoft/sml/sml-std-basis.html}
to ensure a safe and possibly optimized handling of standard data
structures like lists etc. The parser for the \foetus\ terms has been
created with ML-Lex and ML-Yacc.
The ML implementation currently used is
\htmladdnormallink{Standard ML of New Jersey}%
{http://cm.bell-labs.com/cm/cs/what/smlnj/index.html}, Version 109.32.

\begin{table}[htbp]
  \begin{center}
\begin{tabular}{ll}
\srcfile{foetus.lex}   & \foetus\ language token specification for {\tt
  ml-lex} \\
\srcfile{foetus.grm}   & \foetus\ language grammar for {\tt ml-yacc} \\
\srcfile{aux.sml}      & auxiliary functions \\
\srcfile{closure.sml}  & terms and environment \\
{\bf \srcfile{foetus.sml}}  & values, evaluation function {\tt hnf},
  printing \\
\srcfile{matrix.sml}        & polymorphic matrices with necessary operations\\
\srcfile{simpledeps.sml}    & simple implementation of dependecies \\
{\bf \srcfile{analyse.sml}} & static analysis of \foetus\ code \\
{\bf \srcfile{check.sml}}   & termination check via call graph \\
\srcfile{top.sml}      & top level environment \\
\srcfile{load.sml}     & loader and \foetus\ parser
\end{tabular}
    \caption{\foetus\ source files}
    \label{tab:src}
  \end{center}
\end{table}

\download

\section{Conclusion\label{sec:conclusion}}
We have seen that \foetus\ and its ``brain'', the call graph
completion and finding a lexical order on the function arguments,
contributes to automated termination proofs. Of course, in its current
state it is no more than a toy to gather experience on his
subject. Some improvements have to be done: \foetus\ should be able to
recognize more kinds of dependencies (see section \ref{sec:funex}).
\begin{itemize}
\item Let assignments. The use of {\tt let}-constructs to save
  values within functions is discouraged because \foetus\ stores no
  relations concerning them; it performs no symbolic evaluation during
  analyzation. For example:
\begin{verbatim}
case list of
        {Cons pair => let
                hd = pair.HD,
                tl = pair.TL in ...
\end{verbatim}
\foetus\ does not know that {\tt hd} $\less$ {\tt list} and that {\tt
  tl}  $\less$ {\tt list}. At least such simple assignments (for code
shortening) should be handled.
\item Tuple handling. \foetus\ should trace the dependencies not only
  of the whole tuples but also of their components. At the moment you
  cannot define functions with one tuple as parameter instead of
  separate parameters and still expect a termination proof (see
  example \ref{ex:tupelparm}).
\item Function results. The reason that \foetus\ cannot prove
  termination of {\tt div} (see example \ref{ex:div}) is that it does
  not know $x \not=0 \rightarrow (y-x=0 \lor y-x<y)$. But this could
  be shown for the {\tt sub} function by induction and result in a
  dependency \foetus\ could use \cite{BG96}.
\end{itemize}
Furthermore the call graph completion algorithm could be adopted to
prove termination of parameter permuting functions like {\tt zip} (see
example \ref{ex:parmperm}).

If \foetus\ has ``grown older'' in the described manner it could be
``born into'' one of the ``adult'' program verfication systems or
theorem provers like ALF, Isabelle, LEGO or MuTTI {\bf ;-)}.

\begin{rawhtml}
<p align=center>
<a href="http://www.nrlc.org/abortion/wdlb/wdlb1.html">
<img src="../thumb.gif" alt="fetus, 18 weeks old" border=1>
</a>
</p>
\end{rawhtml}

\nocite{Gie97}
\nocite{Sli96}
\nocite{Sli97a}
\nocite{NPS90}
\nocite{TTu97b}
\nocite{Pau91}

\bibliographystyle{myplain}
\bibliography{mybiblio}

\end{document}